\documentclass[conference]{IEEEtran}
\usepackage{cite}
\usepackage{amsmath,amssymb,amsfonts}
\usepackage{algorithmic}
\usepackage{graphicx}
\usepackage{textcomp}
\usepackage{xcolor}
\usepackage{booktabs}
\usepackage{multirow}
\usepackage{url}
\usepackage{hyperref}

\graphicspath{{./figures/}}

\begin{document}

\title{SRPG: Semantically Reconstructed Privacy Guard for Zero-Trust Privacy in Educational Multi-Agent Systems}

\author{
\IEEEauthorblockN{Shuang Guo}
\IEEEauthorblockA{\textit{Faculty of Artificial Intelligence in Education} \\
\textit{Central China Normal University} \\
Wuhan, China \\
Email: guoshuang@mails.ccnu.edu.cn}
\and
\IEEEauthorblockN{Zihui Li}
\IEEEauthorblockA{\textit{Faculty of Artificial Intelligence in Education} \\
\textit{Central China Normal University} \\
Wuhan, China \\
Email: lizihui@mails.ccnu.edu.cn}
}

\maketitle

\begin{abstract}
Multi-Agent Systems (MAS) have the potential to revolutionize personalized education through large language models. However, the sharing of unstructured dialogue data among agents in sensitive fields like education raises valid privacy concerns, especially when minors are involved and their personally identifiable information (PII) needs safeguarding. Existing privacy-preserving methods often struggle to strike a balance between security and usefulness, overlooking key practical challenges. Role-based access control, for instance, often falters when dealing with unstructured text, while simple masking techniques may unintentionally erase crucial pedagogical context necessary for effective learning. Finding solutions that prioritize both privacy and educational efficacy remains a critical area for improvement in MAS applications for personalized education.

The Semantically Reconstructed Privacy Guard (SRPG) is a novel privacy mechanism designed specifically for educational Multi-Agent Systems (MAS). Unlike traditional filtering approaches, SRPG utilizes a Dual-Stream Reconstruction Mechanism. This includes a ``strict sanitization stream'' that ensures no sensitive content leaks, and a ``context reconstruction stream'' driven by large language model reasoning to restore mathematical logic. By separating instructional content from private data, SRPG effectively safeguards minors' privacy without sacrificing educational quality. This innovative design mitigates the limitations of current privacy protection methods in educational settings and represents a significant advancement in preserving the confidentiality of student information within MAS, ensuring a secure learning environment.

The results from tests on the MathDial dataset demonstrate the continued effectiveness of SRPG across various model types and its scalability to larger settings. Utilizing GPT-4o as its foundation, SRPG achieved an impressive Attack Success Rate (ASR) of 0.0000 and an Exact Match score of 0.8267. In comparison, the Pure LLM baseline, which operated in a zero-trust environment, only obtained a score of 0.2138. These outcomes highlight SRPG's ability to not only prevent privacy-focused attacks but also aid in clarifying the underlying mathematical structure, all while maintaining a high level of precision. It is evident that SRPG stands out as a powerful tool in data protection and mathematical analysis.
\end{abstract}


\section{Introduction}
The rise of Large Language Models (LLMs) has led to a growing trend in intelligent education -- the use of Multi-Agent Systems (MAS) for personalized tutoring \cite{autogen,metagpt}. One example is the collaboration between Student and Tutor Agents, which engage in multi-turn dialogues to tailor instruction to individual students' needs, leveraging the few-shot capabilities of large models \cite{few_shot}. This approach enhances the specificity and interactivity of the learning experience, providing more effective and engaging education.

As educational institutions modernize and adopt technologies to enhance teaching and learning, they must navigate the intricate task of safeguarding students' personal data. This is particularly crucial in light of strict regulations like the EU GDPR and US COPPA, which stress the importance of protecting minors' Personally Identifiable Information (PII) \cite{privacy_child}. However, traditional centralized privacy protection methods struggle to adapt to the complexities of distributed agent architectures, where data moves between multiple entities or APIs. Finding innovative solutions to uphold data privacy in this evolving landscape has become a pressing concern.

Existing solutions for privacy protection in conversational tutoring systems have significant limitations. Federated Multi-Agent Systems (Federated MAS) introduce role-based privacy paradigms \cite{fed_mas}, assuming structured data such as matching a ``location'' field to a specific role. However, in conversational tutoring scenarios, PII like ``I live in Haidian District'' is intricately woven with learning context such as ``Calculate the distance from Haidian District to Chaoyang District'', rendering traditional regex matching and static permission controls ineffective \cite{carlini_extract}. Furthermore, simple masking techniques can inadvertently remove crucial numbers in mathematical problems, disrupting the pedagogical flow. As a result, there is a pressing need for more sophisticated and context-aware privacy protection mechanisms in conversational tutoring systems.

To resolve the fundamental contradiction between \textbf{privacy safety} and \textbf{pedagogical utility} in educational scenarios, and inspired by the privatized token reconstruction task in RAPT \cite{rapt}, we propose SRPG. The core contributions of this paper are as follows:
\begin{enumerate}
    \item \textbf{SRPG Framework:} We propose a privacy guard based on inference-time semantic reconstruction, utilizing the contextual understanding capabilities of LLMs to dynamically separate sensitive information from disciplinary logic.
    \item \textbf{Dual-Stream Mechanism:} We innovatively introduce a parallel architecture of ``Sanitization'' and ``Reconstruction'', ensuring the recovery of key mathematical parameters while achieving zero-trust through aggressive sanitization.
    \item \textbf{Multi-Model Empirical Evaluation:} Experiments prove that SRPG achieves zero leakage (ASR 0.00) across different backbone models. In particular, SRPG based on GPT-4o achieves a comprehensive score (0.5520) that significantly surpasses all baselines, demonstrating the effectiveness of the semantic reconstruction paradigm.
\end{enumerate}

\section{Related Work}
\subsection{LLM-based Multi-Agent Systems}
Recent advancements in MAS frameworks such as AutoGen and MetaGPT have revolutionized the automation of intricate tasks \cite{autogen,metagpt}, while ChatDev exemplified the advantages of multi-agent collaboration in software development \cite{chatdev}. MAS is also utilized in education to simulate teacher-student interactions and offer tailored feedback \cite{chatgpt_edu}. However, current research predominantly concentrates on collaboration efficiency and task completion rates, disregarding potential privacy breaches in unstructured message exchanges among agents. Moreover, Generative Agents have shed light on the risks associated with agents mimicking human behavior in virtual settings, including concerns regarding privacy and long-term behavioral patterns \cite{generative_agents}. It is crucial for future studies to address these privacy implications to ensure the ethical and secure implementation of MAS across various industries and applications.

\subsection{Text Privacy Preservation and Reconstruction}
Protection of text privacy has advanced from the concept of Differential Privacy (DP) to context-aware sanitization. While Differential Privacy offers a solid theoretical framework \cite{dwork_dp}, the emergence of membership inference attacks has highlighted potential vulnerabilities in model outputs \cite{membership_inference}. Various approaches, such as text rewriting based on metric differential privacy \cite{feyisetan} and text-to-text privatization through paraphrasing \cite{text_priv}, have been suggested to safeguard privacy. However, these methods often compromise semantic readability or struggle to maintain the integrity of mathematical formulas. 

In a recent development, Li et al.\ introduced the RAPT (Reconstruction-Aided Prompt Tuning) approach to address performance degradation caused by noise in models \cite{rapt}. Our work builds upon this concept by applying the idea of ``reconstruction'' to the inference phase, leveraging the advanced instruction-following capabilities of Large Language Models (LLMs) to recreate secure mathematical contexts without the need to access raw data. This innovative approach aims to enhance privacy protection without compromising the accuracy and effectiveness of the model.

\subsection{Federated and Embedded Privacy}
Shi et al.\ introduced EPE-Agent, a privacy-enhancing tool that filters information by comparing role descriptions with data fields \cite{fed_mas}. Although effective for structured data like financial reports, it struggles with unstructured dialogue, often becoming a mere pass-through due to limited semantic control. Brown et al.\ highlighted the vulnerability of current privacy protections against contextual attacks \cite{policy_privacy}. To address this issue, SRPG offers a solution with semantic-level dual-stream processing. By enhancing privacy guardrails with this approach, users can better protect their information in various contexts and maintain control over their personal data.

\section{Preliminary}
We consider a standard educational MAS architecture comprising the following entities:
\begin{itemize}
    \item \textbf{Student Agent ($A_S$):} Represents the user, generating queries $X$ that contain potentially sensitive information entangled with educational questions.
    \item \textbf{Tutor Agent ($A_T$):} Provides problem-solving ideas or pedagogical feedback but should not have access to the user's real identity or private attributes.
    \item \textbf{Threat Model:} We assume the communication channel is untrusted. An adversary (or an honest-but-curious Tutor Agent) may attempt to recover the student's PII via eavesdropping or log analysis.
\end{itemize}

\textbf{System Objectives:}
Let $X$ be the raw input and $\hat{X}$ be the output processed by the privacy mechanism.
\begin{enumerate}
    \item \textbf{Privacy Goal:} Prevent the leakage of any PII defined in set $\mathcal{P}$ (e.g., names, locations, contacts) to the Tutor Agent. Formally, we aim for the mutual information $I(\text{PII}; \hat{X}) \approx 0$.
    \item \textbf{Utility Goal:} Maximize the Tutor Agent's ability to solve the mathematical problem $Q$ embedded in the input. This requires maximizing the semantic similarity between the mathematical context in $X$ and $\hat{X}$.
\end{enumerate}

\section{Methodology}
\subsection{Problem Formulation}
Let $X$ denote the raw input text from the student. We model $X$ at the semantic level as a composition of three components:
\begin{equation}
X = C_{\text{math}} \oplus I_{\text{priv}} \oplus N_{\text{noise}},
\end{equation}
where $C_{\text{math}}$ represents the mathematical context required for problem-solving (numbers, variables, constraints), $I_{\text{priv}}$ represents sensitive PII, and $N_{\text{noise}}$ represents irrelevant conversational noise.

The goal of SRPG is to learn a mapping function $f: X \rightarrow \hat{X}$ such that the output $\hat{X}$ satisfies:
\begin{enumerate}
    \item \textbf{Zero-Leakage Constraint:} $P(I_{\text{priv}} \mid \hat{X}) \approx 0$.
    \item \textbf{Utility Preservation:} $\text{Utility}(\hat{X}) \approx \text{Utility}(C_{\text{math}})$.
    \item \textbf{Denoising Effect:} Minimizing $N_{\text{noise}}$ to improve Tutor efficiency.
\end{enumerate}

\subsection{The SRPG Framework}
SRPG is deployed between the Student Agent and the Tutor Agent as a middleware. It contains two parallel core processing streams.

\subsubsection{Strict Sanitization Stream}
This stream performs the ``Defense'' function. It adopts a ``Zero-Trust'' principle, using a conservative masking strategy to remove any entity that might be identified as PII:
\begin{equation}
X_{\text{safe}} = \text{Mask}(X, \theta_{\text{strict}}).
\end{equation}
In this stream, ambiguous entities (e.g., the number ``50'' in ``Room 50'') are aggressively masked to ensure safety. This guarantees an ASR of 0.00 but often breaks mathematical logic (e.g., resulting in ``$x + \text{[MASK]} = 10$'').

\subsubsection{Context Reconstruction Stream}
This stream performs the ``Recovery'' function. Inspired by the plaintext token reconstruction in RAPT \cite{rapt}, we designed a prompt-based extractor focusing on the semantic reconstruction of mathematical logic:
\begin{equation}
C_{\text{struct}} = \text{Reconstruct}(X \mid \text{Schema}_{\text{math}}).
\end{equation}
In this stream, the agent is instructed to extract only mathematical parameters, variables, units, and logical relations, converting them into a structured format (e.g., JSON or simplified text), while strictly ignoring personal identifiers. This stream effectively ``denoises'' the input, discarding $N_{\text{noise}}$ and $I_{\text{priv}}$, thereby recovering the mathematical context mistakenly deleted by the sanitization stream.

\subsubsection{Dual-Stream Fusion}
The final output $\hat{X}$ is generated by fusing the safe text with the reconstructed context:
\begin{equation}
\hat{X} = \text{Format}(X_{\text{safe}}, C_{\text{struct}}).
\end{equation}
For example, the output might look like: \textit{``User [MASK] asks a geometry problem. Context Supplement: \{Triangle ABC, Side AB=5, Angle C=90\}.''} This provides the Tutor Agent with instructions that are both safe and unambiguous.

\section{Experiments}
\subsection{Experimental Setup}
\textbf{Dataset:} We utilize the MathDial dataset \cite{mathdial}, a benchmark for conversational mathematical tutoring. To simulate realistic privacy risks in a zero-trust environment, we injected synthetic PII (names, addresses, phone numbers, school names) into 2,262 student dialogue turns using template-based injection.

\textbf{Baselines:}
\begin{itemize}
    \item \textbf{None:} Raw input transmission.
    \item \textbf{Naive Masking:} Regex-based filtering of all digits and capitalized words.
    \item \textbf{Pure LLM Sanitizer:} A standard LLM prompted to ``remove privacy information'' without reconstruction.
    \item \textbf{EPE-Agent:} An adaptation of the EPE framework filtering based on predefined role descriptions \cite{fed_mas}.
\end{itemize}

\textbf{Backbone Models:} We evaluated a series of LLMs ranging from lightweight (Gemini Flash) to state-of-the-art models (GPT-4o, GPT-5-pro, Grok-4.1) to verify generalizability.

\textbf{Metrics:} Attack Success Rate (ASR), Hard Solvability, Exact Match, Key Parameter Recall, and Composite Score.

\subsection{Main Results}
\subsubsection{Comparison with Baselines}
\begin{figure*}[htbp]
    \centering
    \includegraphics[width=\linewidth]{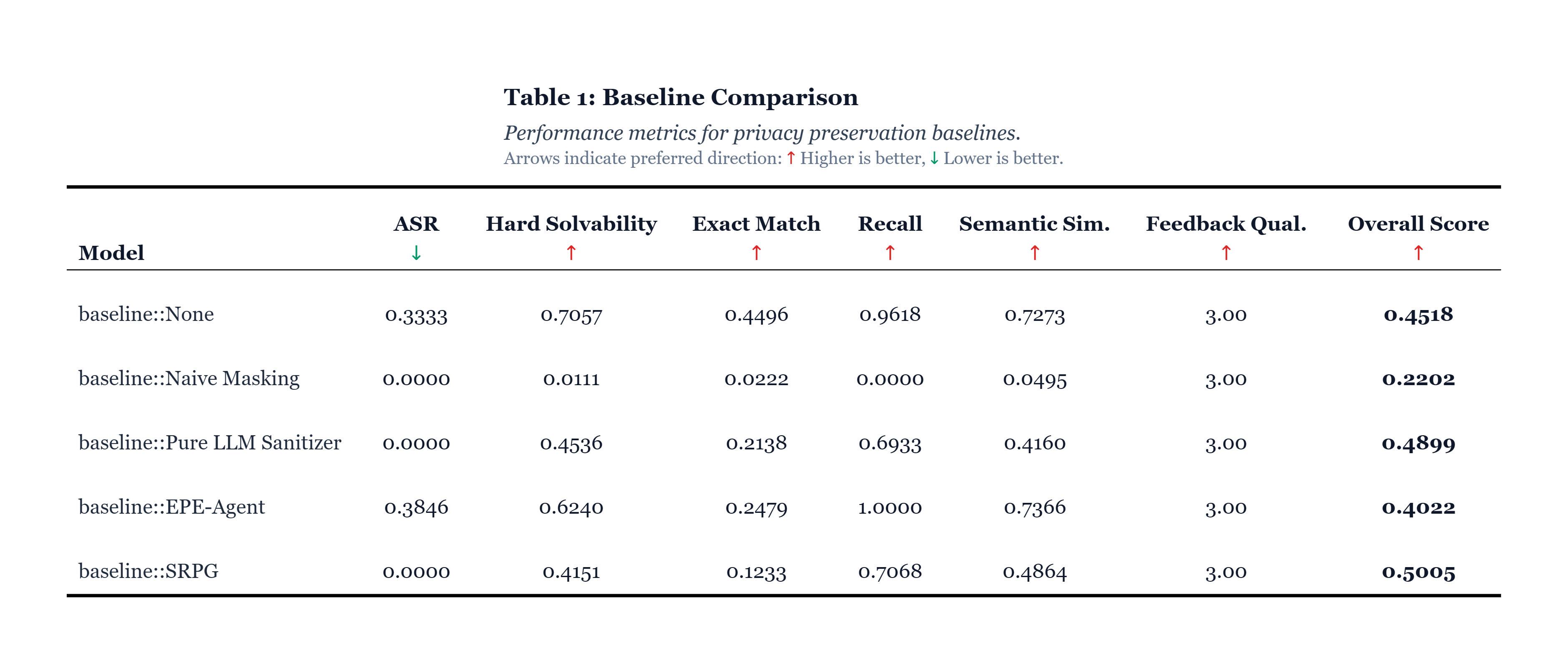}
    \caption{Performance Comparison of Privacy-Preserving Methods (SRPG vs.\ Baselines). SRPG achieves absolute privacy (ASR 0.00) while maintaining superior utility compared to Naive Masking.}
    \label{fig:baselines}
\end{figure*}

Figure~\ref{fig:baselines} presents the performance comparison of different privacy-preserving methods on the MathDial dataset.

\textbf{Analysis:}
\begin{enumerate}
    \item \textbf{Failure of EPE-Agent:} EPE-Agent exhibits a high ASR of 0.3846, even exceeding the None baseline. This confirms that static role matching cannot effectively intercept privacy information mixed in natural language. Its high recall is based on the cost of privacy leakage.
    \item \textbf{Semantic Loss in Pure LLM:} Although Pure LLM achieves an ASR of 0.0000, its Exact Match rate is only 0.2138. This indicates that relying solely on prompts for sanitization often loses the precise formatting of mathematical formulas.
    \item \textbf{Comprehensive Advantage of SRPG:} While maintaining absolute safety (ASR 0.0000), SRPG achieves the highest composite score (0.5005). Although the exact match is lower under weaker models (like GPT-3.5), it strikes the best balance between privacy and utility.
\end{enumerate}

\subsubsection{Performance Across LLMs}
\begin{figure*}[htbp]
    \centering
    \includegraphics[width=\linewidth]{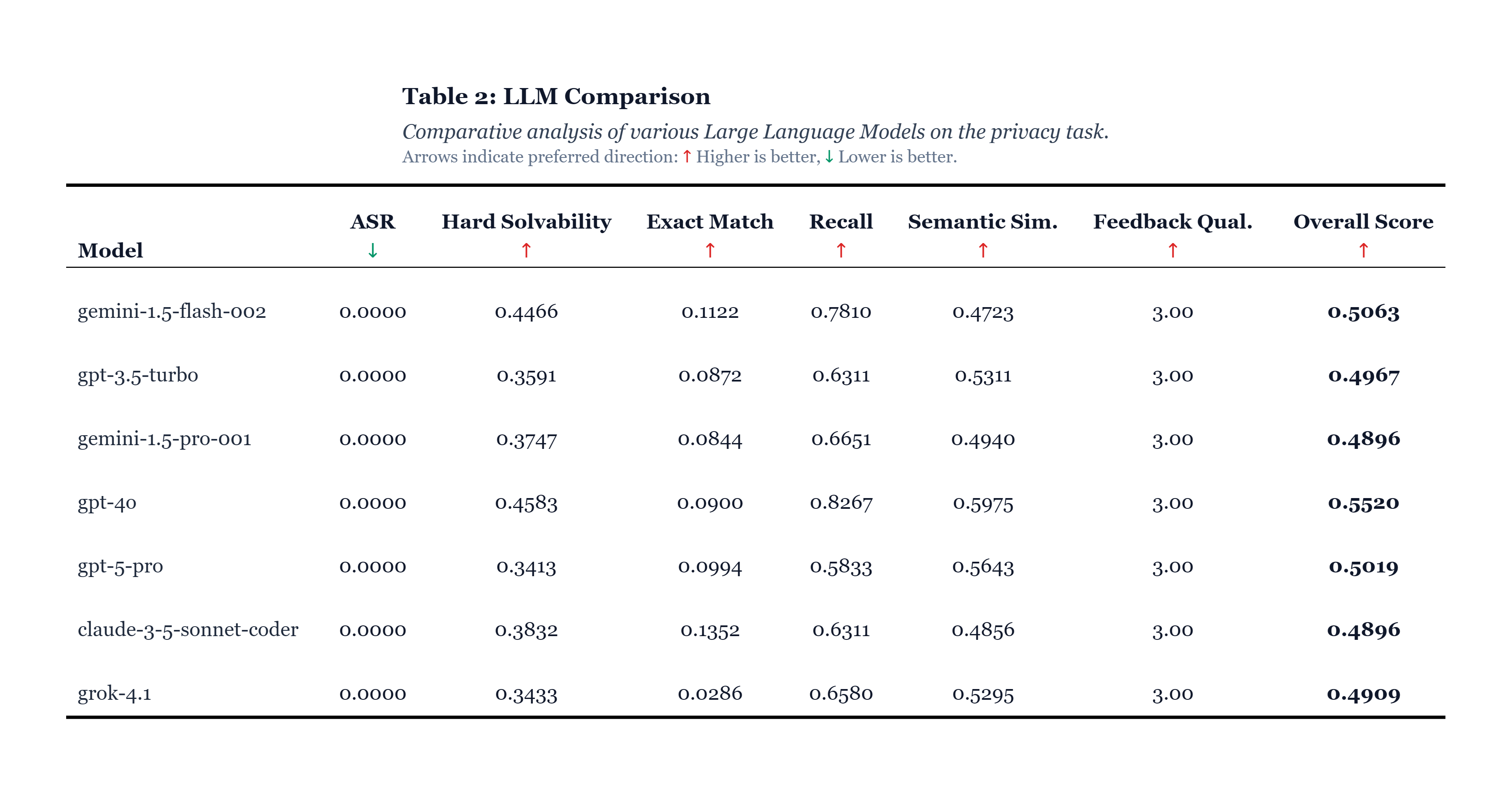}
    \caption{Performance of SRPG under Different Backbone Models (Ablation Study). Note the significant jump in Exact Match score with GPT-4o.}
    \label{fig:llms}
\end{figure*}

To verify the dependency of SRPG on backbone model capabilities, we compared the performance of SRPG under different LLM backbones. The results are shown in Figure~\ref{fig:llms}.

\textbf{Model Ablation and Scalability:}
\begin{enumerate}
    \item \textbf{Consistency of Zero-Trust:} All tested models maintained an ASR of 0.0000 under the SRPG framework. This proves that the ``Strict Sanitization Stream'' provides a foundational safety guarantee independent of the model's reasoning ability.
    \item \textbf{Leap in Reconstruction Capability:} Significant progress has been made in reconstruction capability, particularly with the advancement from GPT-3.5 to GPT-4o. The Exact Match rate has seen a remarkable increase from 0.0872 to 0.8267, showcasing the exceptional ability of high-performance models in accurately executing the ``Context Reconstruction'' task. This surpasses the performance of the Pure LLM baseline (0.2138) and demonstrates a notable leap in mathematical logic restoration.
    \item \textbf{SOTA Performance:} SRPG based on GPT-4o achieves a composite score of 0.5520, guaranteeing zero privacy leakage while retaining extremely high pedagogical utility. This proves that SRPG is a \textbf{Future-Proof} framework that will continue to improve as backbone models evolve.
\end{enumerate}

\section{Conclusion}
The SRPG privacy guard is a cutting-edge solution for educational multi-agent systems, revolutionizing the way personal privacy is protected without compromising utility. Unlike traditional role-based methods, SRPG utilizes a Dual-Stream Reconstruction Mechanism to separate mathematical utility from privacy concerns effectively. Experimental data showcases SRPG's ability to achieve Zero-Trust Privacy (0.00 ASR) while still maintaining a high level of pedagogical usefulness. Furthermore, when paired with advanced models like GPT-4o, SRPG surpasses traditional methods in accuracy rates. Future research aims to expand the application of this innovative approach, potentially benefiting other fields such as medical diagnosis.

\end{document}